\documentclass[%
 reprint,
 amsmath,amssymb,
 aps,
]{revtex4-2}

\usepackage{graphicx}
\usepackage{dcolumn}
\usepackage{bm}
\usepackage{xcolor}

\begin{document}

\preprint{APS/123-QED}

\title{Semi-Markov random walk on complex networks}

\author{Lasko Basnarkov}
 \email{lasko.basnarkov@finki.ukim.mk}
\affiliation{SS. Cyril and Methodius University, Faculty of Computer Science and Engineering, Skopje, Macedonia}
 \affiliation{Research Center for Computer Science and Information Technologies, Macedonian Academy of Sciences and Arts, Skopje, Macedonia}

\date{\today}

\begin{abstract}
We present a semi-Markov model of random walk on complex networks in discrete and continuous-time scenario. In the general setting of the semi-Markov chains, the duration of stay at given node -- the sojourn time -- is random, and the probability to transition to a neighbor depends on the sojourn time. Analytical formulae for the average sojourn time and the node occupation probability of infinite walk are presented and verified with Monte Carlo simulations for two examples. We propose an application of the semi-Markovian random walk for ranking web pages determined by the fraction of time that infinite random surfer spends on a web page -- time rank, as an alternative to the existing PageRank that relies on the fraction of visits -- visit rank. 
\end{abstract}

\maketitle



{\it Introduction} -- Random walk is a mathematical concept that has found application in description of phenomena nearly across all scientific disciplines. The Brownian motion of suspended particles \cite{wang1945theory}, irregularities in the price fluctuations in the stock market \cite{bachelier1900theorie}, ranking of web pages \cite{brin1998anatomy}, and practical sampling from arbitrary probability distribution \cite{gilks1995markov} are just few examples showing the diversity of the implementation of the Markov chains and processes as models for mathematical description of the random walk which is considered as basic mechanism in those phenomena. In the physics community, it has been noticed vivid activity in studying the effects of random searching in non-trivial structures like complex \cite{noh2004random} and multiplex \cite{de2013random} networks, quantifying node centralities \cite{newman2005measure, borgatti2005centrality}, design of immunization strategies \cite{holme2004efficient}, while the most recent wave of interest in the area is the analysis of the random walk with resetting \cite{evans2011diffusion}. 

The design of the various models of random walk relies on the very plausible Markovian assumption -- the probability of any future outcome depends only on the current state, and not on the previous ones. Furthermore, the  Markov models also assume: 1. Exponential distribution of the sojourn time -- the time in which the system resides in given state before transition to another; and 2. Time-homogeneity -- the transition probabilities to other states are constants. These principles appear as gold standards in the various models of random walk. In a more general setting, one applies renewal theory, which considers that the distribution of sojourn time is an arbitrary, not necessarily exponential function. The generalization can be made one step further by assuming that the probabilities to transition to the other states depend on the sojourn time, which means that one has a semi-Markov model \cite{korolyuk1975semi}. It is interesting  that besides being introduced around the mid of the past century \cite{takacs1954some,smith1955regenerative,levy1956semi}, semi-Markov models are not so popular as the simpler counterparts are. Examples of their use can be found in reliability theory \cite{barbu2009semi}, finance \cite{janssen2007semi}, quantum mechanics \cite{breuer2008quantum}, machine learning \cite{yu2010hidden}, and so on. 

In this Letter we will study random walk on complex networks, where the jumping probabilities to the other nodes depend on the time that has lapsed since the walker has entered in a given state, and the sojourn time is not necessarily exponentially distributed. We present an analytical formula for the mean occupation time of the nodes by an infinite walk and apply it to rank the nodes according to this quantity. Its potential application is the use of this value to rank the web pages with its interpretation as an estimate of the expected time that random surfer spends on given page. 



{\it Discrete-time model} -- Consider a discrete-time random walk on a complex network with $N$ nodes. Without losing generality, assume that the walker has entered at node $i$ at moment 0, and since in tne Markovian setting the probabilities do not depend on the past nodes, anytime a jump is made on a given node the clock is reset to zero. Denote with $\tau$ the local time, that is the time that has lapsed since the walker has entered certain node $i$, which we will call {\it age of state}, or just {\it age}. The respective survival function (the probability that the walker is still at node $i$ at age $\tau$) is $S_i(\tau)$ . Let the conditional probability to jump to a neighbor $j$ at age $\tau$ is $p_{ij}(\tau)$, with $p_{ii}(\tau)$ being the probability to remain at the same node, and thus one has the normalization condition $\sum_j p_{ij}(\tau) =1$, for any $\tau$. As a side note, if these probabilities are constants $p_{ij}(\tau) = p_{ij}$, one has ordinary Markov chain. The survival function can thus be expressed through the conditional probabilities as
\begin{equation}
    S_i(\tau) = p_{ii}(0)p_{ii}(1) \dots p_{ii}(\tau). \label{eq:survival}
\end{equation} 
Now, the probability that the walker will jump from node $i$ to $j$ after being $\tau$ time steps in $i$ is
\begin{equation}
    P_{ij}(\tau) = S_i(\tau-1)p_{ij}(\tau).\label{eq:jump}
\end{equation}
We further assume that immediate transition is impossible, so $p_{ii}(0) = 1, p_{ij}(0) = 0$, and that for each node there is some maximal possible sojourn time $T_i$, so that $p_{ii}(T_i)=0$. For this kind of random walk, one can construct a related Markov chain, in which the state is the pair $(i, \tau)$ identified with the state in the original semi-Markov chain (the node $i$) and age $\tau$. If all $N$ nodes are communicating -- any node $i$ can be reached from another $j$, the related Markov chain is ergodic. Respectively, there exists a stationary distribution $\pi_i(\tau)$ for each such state $(i, \tau)$. By using survival relationship (\ref{eq:survival}), one has that for each $i$ the stationary probabilities are related with 
\begin{eqnarray}
    \pi_i(\tau) &=& \pi_i(\tau - 1) p_{ii}(\tau) = \pi_i(0)p_{ii}(1)\dots p_{ii}(\tau) \nonumber\\ &=&\pi_i(0)S_i(\tau).
\end{eqnarray}
In order to find the stationary probabilities $\pi_i(\tau)$, from the conditional probabilities $p_{ji}(\tau)$, we can consider the event of jump to new node $i$, from all other nodes $j$, and from any possible sojourn time in those nodes. Then, the stationary probability $\pi_i(0)$ corresponding to that event is
\begin{eqnarray}
    \pi_i(0) &=& \sum_{j=1}^{N} \sum_{\tau=1}^{T_j-1}\pi_j(\tau-1)P_{ji}(\tau)\\ 
    &=&\sum_{j=1}^{N} \pi_j(0) \sum_{\tau=1}^{T_j-1}S_j(\tau-1)p_{ji}(\tau) = \sum_{j=1}^{N} \pi_j(0) \Pi_{ji},\nonumber
\end{eqnarray}
where we have denoted the transition parameter from state $j$ to $i$ as
\begin{equation}
    \Pi_{ji} = \sum_{\tau=1}^{T_j-1}S_j(\tau-1)p_{ji}(\tau). \label{eq:Trans_prob_disc}
\end{equation}
Thus we have obtained that the vector of probabilities $\pi_i(0)$ is the left principal eigenvector of the transition matrix $\Pi$ which has elements $\Pi_{ji}$. In general, this matrix is not stochastic and its elements are not probabilities and can even have values larger than one. Note also that $\pi_i(0)$ is not a probability vector, since $\sum_i \pi_i(0) < 1$, so appropriate normalization is needed, which will be provided below. 

An important quantity is the mean sojourn time, that represents the expected number of steps in which the walker is at given node $i$ before jumping to another one
\begin{equation}
    w_i = \sum_{\tau=0}^{T_i-1} \tau S_i(\tau).  
\end{equation}
The more important quantity we are interested in is the fraction of time in which the walker is at given node. Since all stationary probabilities sum up as $\sum_{i, \tau} \pi_i(\tau) = 1$, we have the occupation probability $\pi_i$ -- the probability to be in state $i$ (with any possible age $\tau)$, which is also the fraction of time the walker in infinite random walk will be at that node, given with
\begin{equation}
    \pi_i = \sum_{\tau = 0}^{T_i - 1} \pi_i(\tau) = \pi_i(0) \sum_{\tau = 0}^{T_i - 1} S_i(\tau).
\end{equation}
An approach to find the mean sojourn time and the occupation probability, is to take an arbitrarily scaled principal left eigenvector $\mathbf{u}$ of the transition matrix $\Pi$ and take that $\pi_i(0) = a u_i$, for some constant $a$. Then the constant $a$ would be found from the normalization condition
\begin{equation}
    1 = \sum_{i,\tau} \pi_i(\tau) = \sum_i \pi_i(0) \sum_{\tau = 0}^{T_i - 1} S_i(\tau) = a \sum_i u_i \sum_{\tau = 0}^{T_i - 1} S_i(\tau). \label{eq:Norm_const}
\end{equation}


{\it Continuous-time model} -- For completeness,  consider also random walk in a network with $N$ nodes where the time between jumps can have a random value drawn from continuous distribution. Let the probability to jump from $i$ to another node at age $\tau$ is determined with time-dependent exit rate $\lambda_i(\tau)$. Then the exit probability at age $\tau$ is given with
\begin{equation}
    \frac{dS_i}{d\tau} = -S_i(\tau)\lambda_i(\tau),
\end{equation}
where $S_i(\tau)$ is the respective survival function. The last relationship can be solved to find the survival function as
\begin{equation}
    S_i(\tau) = e^{-\int_{0}^{\tau} \lambda_i(t)dt}.
\end{equation}
Note that for continuous-time (ordinary) Markov chain one has constant rate $\lambda_i(\tau) = \lambda_i$, and respectively exponential survival function $S_i(\tau) = e^{-\lambda_i\tau}$. For the semi-Markovian case, the transition probabilities and the transition rates are time-dependent $\lambda_{ij}(\tau)$. So, the probability for transition from node $i$ to node $j$ within the interval $(\tau, \tau+d\tau)$ is
\begin{equation}
    p_{ij}(\tau)d\tau = S_i(\tau)\lambda_{ij}(\tau)d\tau.
\end{equation}
We note that due to the conservation of the probability, one has the following relationship between rates
\begin{equation}
    \lambda_i(\tau) = \sum_{j=1}^{N} \lambda_{ij}(\tau). \label{eq:lambda_conservation}
\end{equation}
Denote with $\pi_i(t)$ the probability density corresponding to the event of jumping to the node $i$ which has happened in the interval $(t, t+dt)$. So, similarly to the discrete-time case we will have a relationship 
\begin{eqnarray}
    \pi_i(t) dt &=& \left(\sum_{j=1}^{N} \int_{0}^{\infty}\pi_j(t-\tau)p_{ji}(\tau)d\tau\right) dt \\ 
    &=&\left(\sum_{j=1}^{N} \int_{0}^{\infty}\pi_j(t-\tau) S_j(\tau) \lambda_{ji}(\tau)d\tau\right) dt,\nonumber
\end{eqnarray}
where $\pi_j(t-\tau)$ accounts for the jumps to the nodes $j$ at moment $t-\tau$ in the past from where transition to $i$ is made at moment $t$. If the network is strongly connected, then by similar reasoning as in the discrete-time case, one can conclude that there is a stationary distribution. This will result in time-invariant probability density $\pi_i(t) = \pi_i$ for each node, which leads to
\begin{equation}
    \pi_i = \sum_{j=1}^{N} \pi_j \int_{0}^{\infty}S_j(\tau)\lambda_{ji}(\tau)d\tau = \sum_{j=1}^{N} \pi_j \Pi_{ji},
\end{equation}
where the continuous-time transition parameters are
\begin{equation}
    \Pi_{ji} = \int_{0}^{\infty}S_j(\tau)\lambda_{ji}(\tau)d\tau.
\end{equation}

Thus, we have obtained similar relationship as the one for the discrete-time case (\ref{eq:Trans_prob_disc}). Note that for the exponential survival time $S_j(t) = e^{-\lambda_j t}$ (and respectively constant $\lambda_j(t) = \lambda_j$), one has
\begin{equation}
    \Pi_{ji} = \frac{\lambda_{ji}}{\lambda_j},
\end{equation}
which represent the transition probabilities in a continuous-time Markov chain.

{\it Examples} -- Consider a network of $N$ web pages, with hyperlinks from one to another page and assume for simplicity that the amount of the text (or other content) at each page is proportional to the number of hyperlinks within that page, that is outgoing links in network terminology. Assume that a random surfer would spend one time unit at that page in reading the content from one to the next link. Upon reaching a link she, or he decides either to click on the link with certain probability or continue reading. We assume that the surfer must click on the last link, so we have a surfing without stopping. To describe this scenario we need a complex network of $N$ nodes that represent the pages and links between the nodes that correspond to the hyperlinks. Thus the maximal sojourn time for each node $i$ is $T_i = k_i$, where $k_i$ is its (out-)degree. We order the neighbors of the node as the respective hyperlinks appear, so under the proposed reading scenario at age $\tau$, the random walker in the network can stay at the same node (proceed with reading) with probability $p_{ii}(\tau)$, or click on the link with probability $p_{ij}(\tau)$ -- go to the page $j$, which according to the ordering is $\tau$-th neighbor. Note that, at age $\tau$, $p_{ik}(\tau) = 0$, for any other  neighbor $k$, so $p_{ii}(\tau) + p_{ij}(\tau) = 1$. Assume for simplicity that the chance to proceed to any neighbor of $i$ is equal. To achieve that, one has to take that at age $\tau = 1$, the probability to jump to the first neighbor is $1 / k_i$, while the probability to stay in the same node is $1 - 1 / k_i$. Next, at age $\tau = 2$, take that the walker can jump to the second neighbor with probability $1 / (k_i-1)$ which gives the probability to jump to the second page as $(1 - 1 / k_i) / (k_i-1) = 1/k_i$. The probability to stay at the same node at age $\tau=2$ is obtained from the normalization condition. Thus, at arbitrary age $\tau$ the probability to jump to the $\tau$-th neighbor is $1 / (k_i-\tau + 1)$, so the walker can jump to each neighboring node with equal probability $1/k_i$, but at different age. By using these values of the transition probabilities, one can find the left principal eigenvector of the matrix $\Pi$ and use the normalization relationship (\ref{eq:Norm_const}) to determine the occupation probabilities.

\begin{figure}[htb]
\centering
\vspace{0mm}
\includegraphics[width=0.45\textwidth]{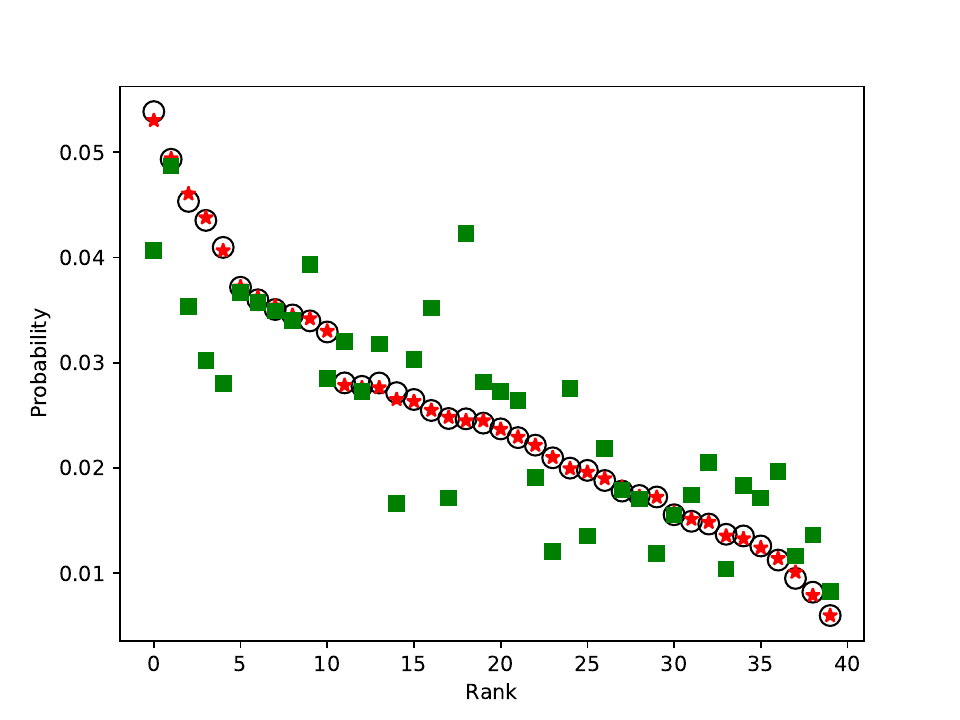}\caption{\label{fig:HighEn} Theoretical estimates of the occupation probability (red stars), compared to that from the Monte Carlo simulations (black circles), and the classical random walk (green squares). The nodes are ordered by the rank based on the occupation probability by semi-Markov random walk. The results are for random network with $N=40$ nodes with each link between nodes appearing with a probability of 0.15.
\vspace{0mm}}
\label{fig:TheorSim}
\end{figure}

To verify the theoretical predictions we have generated a directed Erd\"{o}s-R{\'e}nyi \cite{Erdos1959RG} complex network with $N=40$ nodes by randomly generating links between the nodes with probability 0.15. The order of the links, that should represent the order of appearance of the hyperlinks of the hypothetical pages was chosen randomly. Besides the theoretical predictions of the mean occupation time we have made a numerical simulation of the random walk and calculated the average occupation time for each node. The values from numerical simulations are obtained from averaging over a period of $10^6$ time units, that has followed a transition period with the same length. The nice accordance between the theoretical predictions and the simulations can be noticed in the figure \ref{fig:TheorSim}. For comparison, we have also determined the classical ranking of the nodes, where the probability to jump to each neighbor is equal but the sojourn time is one. The redistribution of the ranks resulting from changes in the respective occupation probabilities is evident. We note that besides providing alternative ranking, the semi-Markovian approach offers practical application by relying on the estimate of the average time the surfer spends on a given page, which could be relevant, for example, for advertisement purposes.

\begin{figure}[htb]
\centering
\vspace{0mm}
\includegraphics[width=0.45\textwidth]{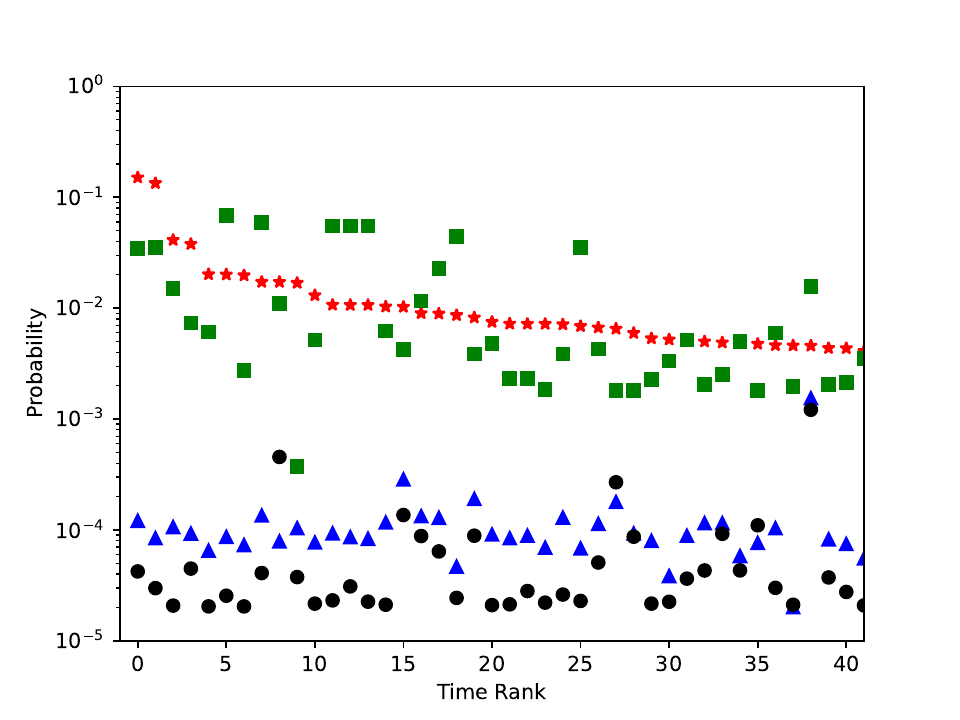}\caption{\label{fig:HighEn} Comparison of the occupation probabilities for the High Energy Physics Theory citation dataset. The meaning of the symbols is the following: red stars -- semi-Markov random walk; green squares -- Markov random walk; black circles -- PageRank algorithm with semi-Markov random walk; blue triangles --  PageRank algorithm with Markov random walk. The horizontal axis represents the time rank of given node. Note the logarithmic scale of the vertical axis. 
\vspace{0mm}}
\label{fig:HEPT}
\end{figure}

As a real world example, instead of looking at a dataset with a network of web pages, which are in general very large, and consequently need large computation capabilities, we have opted to use the High Energy Physics Theory citation network -- a graph made of papers in the respective field with links representing the citations \cite{leskovec2005graphs}. The dataset represents a directed network, which is not strongly connected, which means that there are pairs of nodes such that there is no path between them. Thus, our analysis was made on the largest strongly connected component which has 7464 nodes. We have used the same semi-Markov model as in the previous example, so that the maximal sojourn time equals the number of outgoing links, and the transition probabilities to all neighbors to which there is a link are equal to each other. Since the position of the citation of other papers is absent in the dataset, we have made ordering of the links based on circular permutation of the order of the nodes in the dataset. In this way as first neighbor (first hyperlink) of the node $i$ was considered the first node $j$ toward which $i$ has a link and such that $j$ is the smallest number such that $j>i$, or $j+N>i$. The same reasoning was applied to determine the second link, the third one, and so on. The interpretation of the occupation probability for this case can be the mean time that a typical reader of papers in this field (and which does not stop reading) would spend on the given paper, by assuming that the length of paper is proportional to the number of papers from this dataset, that are cited in this paper. In the figure \ref{fig:HEPT} are given the occupation probabilities of the nodes for the semi-Markovian random walk, along with the classical random walk. The horizontal axis represents the node rank by the occupation probability of the semi-Markov random walk, so we have presented the results for the top 40 nodes. It can be noticed that with this kind of random walk there is a sharp increase in occupation probability when one goes towards the top nodes, as compared to the more smooth increase in the case of the classical random walk. It should be emphasized that one could work with larger networks and find the principal eigenvector by raising the transition matrix to some large power, but only when the matrix is stochastic, which is restricted to the choice of the transition probabilities $p_{ij}(\tau)$. For general case this will not work, and the occupation probability can be found only by finding the principal eigenvector directly, which limits the size of networks which can be practically considered. If one looks at the number of neighbors of the highly ranked nodes by the proposed ranking it will be obtained that they have many incoming links which is sign of frequent visits, but also many outgoing links, which corresponds to longer sojourn time. As it is well known, a node will have nigher rank if the incoming links are from other highly ranked nodes.

The PageRank algorithm \cite{brin1998anatomy} assumes that the surfer would continue surfing by  randomly clicking on the hyperlinks with certain probability $d$, which is also known as damping factor, or decide to jump to arbitrary page with probability $1-d$. To make a generalization of this scenario within our framework, we apply similar reasoning as in the original algorithm. At each time step the probability to follow the semi-Markov random walk is $d=0.85$, which is the standard value of the damping factor, and with probability $1-d$ the jump is made to a node that is selected with probability $1/N$. Thus, the conditional transition probability has value $(1-d)/N$ for each node that is not neighbor to the current node $i$, while for a neighbor $j$ it is a combination of random jump and the random walk $(1-d)/N + dp_{ij}(\tau)$. The conditional probability to remain in the same node for one more step is  $dp_{ii}(\tau)$, while to stay at the same node, but to reset the clock is $(1-d)/N$, like jumping to a non-neighbor. This implies also that the survival function for the PageRank algorithm is $S_{i,PR}(\tau) = d^{\tau}S_i(\tau)$, where $S_i(\tau)$ is the survival function for the case without restarts (for $d=1$). The resulting values of the transition parameters (\ref{eq:Trans_prob_disc}) are now straightforward. In the figure \ref{fig:HEPT} we have also added for comparison the results of the PageRank algorithm with the semi-Markovian random walk and the classical random walk. One can notice that the occupation probabilities corresponding to the top ranked nodes from the semi-Markovian random walk are much smaller in this case. This can be attributed to the resetting which results in jumping elsewhere in the network.
It should be noted that the PageRank algorithm with Markovian random walk can be applied to a network that is not strongly connected, and the principal eigenvector can be obtained from the powers of the transition matrix $\Pi$, which is doable for much larger matrices than we have used here. However, for the semi-Markovian setting, the transition matrix $\Pi$ would not be a stochastic like in this example, so the matrix power method cannot be applied. 

{\it Discussion} -- The semi-Markov models are the most general form of random processes without memory, and which allow for closed form expression for certain quantities like the occupation probability and mean sojourn time. When applied to describe random walk on a complex network they can provide a more realistic model of a process that is going on the network. The resulting occupation probability of infinite random walk can be used for ordering of nodes which puts the focus on the average time of presence in a node and can be called {\it time ranking} as compared to the classical random walk that accounts for the expected number of visits -- {\it visit ranking}. The time ranking is thus providing different perspective, but might be also more appropriate quantification of the importance of the nodes in certain applications. The specific model that was proposed here should be refined if one is willing to implement it for ranking web pages, since a page with more hyperlinks does not necessarily have more content, and if the page is longer, it does not mean that its content is more interesting to keep the attention of the page visitors.

There are numerous other aspects of the semi-Markovian models of processes on complex networks that can be further examined. Determination of the Mean First Passage Time as a quantity for estimation of the random searching of target node by a semi-Markov random walker is among the first things that could be tried. Designing efficient search algorithms in this setting is another challenge that has yet to be explored. The influence of considering more general form of random movement without memory in regular structures like a plane, or regular lattices might provide further insight of this more general form of random walk. The ever increasing abundance of data might be a chance to check whether these models are relevant and necessary to use for explaining real world phenomena, or the simpler, ordinary Markov models are sufficient.

{\it Acknowledgments} -- This work has been supported by the Faculty of Computer Science and Engineering, from the SS Cyril and Methodius University in Skopje, Macedonia.


\bibliography{apssamp}

\end{document}